\documentstyle[psfig]{caosp}
\begin{document}
\pubyear{1998}
\volume{27}
\firstpage{347}
\htitle{RV studies of roAp stars}

\hauthor{A.P. Hatzes and A. Kanaan }

\title{Radial Velocity Studies of roAp Stars}
\author{A.P. Hatzes \inst{1} and A. Kanaan \inst{2} }
\title{Radial Velocity Studies of roAp Stars}

\institute{McDonald Observatory, University of Texas, USA \and
Departamento de Matem\'atica, UFSM, Brazil}
\maketitle
\begin{abstract}
We present recent results from  our radial velocity (RV)
 studies of rapidly oscillating
Ap (roAp) stars. Our measurements reveal that the pulsational amplitude
of these stars 
depends on the spectral region that is examined. For one star,
$\gamma$ Equ, the pulsational RV amplitude
depends on both line strength and atomic species. The elemental
difference is most likely related to the abundance patches on the 
surface of these stars. The line strength
difference is interpreted
as arising form a height effect in the atmosphere. Ultimately,
these measurements
may provide valuable diagnostics of the atmospheres of these stars.
\keywords{Stars: oscillations - Techniques: radial velocities}
\end{abstract}

\section{Introduction}
	The rapidly oscillating Ap stars (roAp)  represent a class
of stars pulsating in nonradial p-modes with periods of 4--15 min.
We have begun a programme of using precise stellar radial velocity (RV)
measurements to study these pulsations. 
Data were acquired using the F3 focus of the 
2-d coud{\'e} spectrograph of McDonald Observatory's
2.7-m telescope ($R$ $\approx$ 60,000).
RV measurements were made using an iodine
gas absorption cell placed in the optical path of the spectrograph
(Cochran \& Hatzes 1994). 
Exposures of 45-60 secs duration were made in rapid
succession over  $\sim$ 4 hrs for each of the stars presented here.

\section{Results}
\begin{figure}[hbtp]
\centering{
\vbox{\psfig{figure=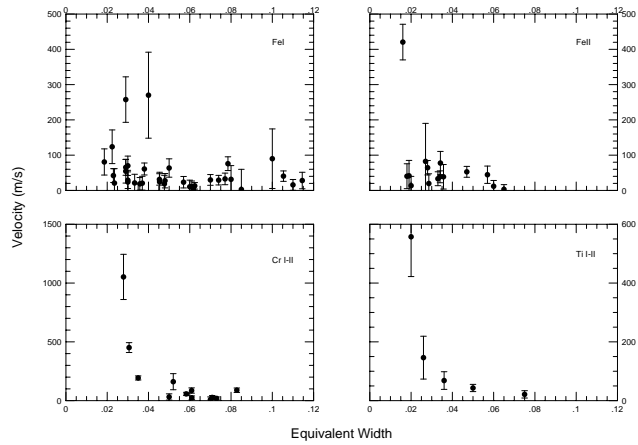,width=9cm,%
    bbllx=0.0cm,bblly=2.0cm,bburx=26.5cm,bbury=22.0cm,clip=
}}\par
       }
\caption{The amplitudes of all unblended lines of Cr\,{\sc i} and {\sc ii},
Ti\,{\sc i} and {\sc ii}, as well as Fe\,{\sc i} and {\sc ii}. The RV 
semi-amplitude is shown as a function of equivalent width.}
\label{gamequ}
\end{figure}

	To date we have measured RVs in three roAp stars: 

{\it $\gamma$ Equ:} The photometric variability is dominated by
4 modes centered around 12 min (Martinez et al. 1996).
The dominant photometric period has been detected in our RV measurements.
A detailed  analysis
(Kanaan \& Hatzes, 1998) reveals  that the pulsational RV
amplitude depended not only on atomic species, but also line strength.
Figure~\ref{gamequ} shows the measured pulsational amplitude as
a function of equivalent width for neutral and once ionized
Cr, Fe, and Ti.
	
\begin{figure}[hbtp]
\centering{
\vbox{\psfig{figure=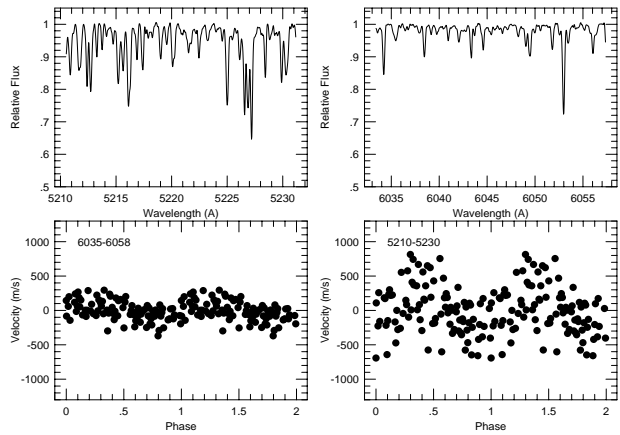,width=9cm,%
    bbllx=0.0cm,bblly=2.0cm,bburx=26.0cm,bbury=22.0cm,clip=}}\par
       }
\caption{The top two panels show two spectral regions (orders) 
of HR 1217. Below each spectrum  is the RV measurements 
from these spectral regions phased to the
photometric period.}
\label{hr1217}
\end{figure}

{\it HR 1217:} There are at least 6 pulsation modes in this 
star with periods near
6.15 minutes. Matthews et al. (1988) found a peak-to-peak
RV amplitude of 400 m s$^{-1}$. Our measurements show that the
RV amplitude depends on the spectral region that is examined 
(Fig.~\ref{hr1217}).

\begin{figure}
\vbox{\psfig{figure=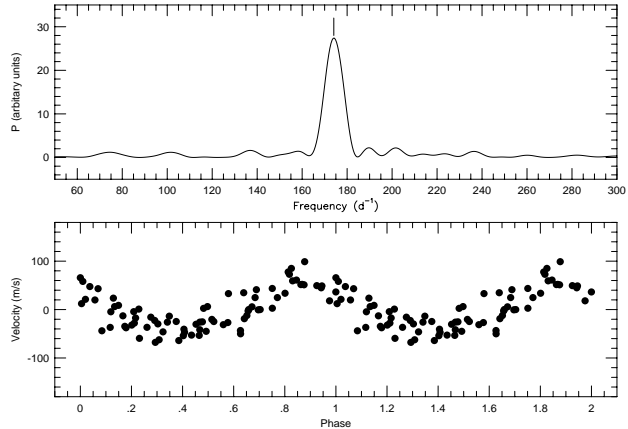,width=9cm,%
    bbllx=0.0cm,bblly=2.0cm,bburx=26.0cm,bbury=22.0cm,clip=}}\par
\caption{(Top) The Scargle-Lomb periodogram of the RV measurements
derived from all spectral lines in the wavelength range 5500 -- 5684 {\AA}.
The vertical line marks the frequency of the photometric period from
Kurtz (1991). (Bottom) The RV measurements phased to the photometric
period.}
\label{33lib}
\end{figure}

{\it 33 Lib:} The single pulsation mode with period 
8.27 min (Kurtz 1991) is clearly seen in 
the periodogram of the RV measurements (top panel of Fig.~\ref{33lib}).
The 2$K$-amplitude of the variations is about 100 m s$^{-1}$ 
(lower panel Fig.~\ref{33lib}).

\section{Discussion}
	Our RV measurements show that the pulsational amplitude
for roAp stars depends 1) on atomic species (e.g. in $\gamma$ Equ,
Cr has a higher amplitude than Fe), and 2) line strengths (weaker
lines have a higher RV amplitude than stronger lines). The former
effect is interpreted as arising from the inhomgenous distribution
of elements on the stellar surface (if Cr is concentrated at the magnetic
poles then its vertical motion is unhindered by the field lines).
The line strength effect most likely arises from vertical structure
of the pulsations since weaker lines are formed, on average, deeper
in the stellar atmosphere.


\end{document}